\documentclass[runningheads]{llncs}
\usepackage{graphicx}

\usepackage{booktabs}
\usepackage{color}
\usepackage{amsmath,amssymb}
\usepackage{mathrsfs}
\usepackage{marvosym}
\usepackage[colorlinks,
            linkcolor=blue,
            anchorcolor=blue,
            citecolor=blue]{hyperref}
\begin{document}
\title{Multi-phase Liver Tumor Segmentation with Spatial Aggregation and Uncertain Region Inpainting}
\titlerunning{Multi-phase Liver Tumor Segmentation}
%
\author{
Yue Zhang\inst{1}$^{(\scriptsize\textrm{\Letter})}$ \and
Chengtao Peng\inst{2} \and
Liying Peng\inst{1} \and 
Huimin Huang\inst{1} \and
Ruofeng Tong\inst{1,3} \and
Lanfen Lin\inst{1} \and
Jingsong Li\inst{3} \and 
Yen-Wei Chen\inst{4} \and
Qingqing Chen\inst{5} \and
Hongjie Hu\inst{5} \and 
Zhiyi Peng\inst{6}}

\authorrunning{Y. Zhang et al.}
%

\institute{College of Computer Science and Technology, Zhejiang University, Hangzhou, China \\
\email{yuezhang95@zju.edu.cn}
\and Department of Electronic Engineering and Information Science, University of Science and Technology of China, Hefei, China  \and
Research Center for Healthcare Data Science, Zhejiang Lab, Hangzhou, China \and
College of Information Science and Engineering, Ritsumeikan University, Kusatsu, Japan \and
Department of Radiology, Sir Run Run Shaw Hospital, Hangzhou, China \and 
Department  of  Radiology, The  First Affiliated  Hospital,  College  of  Medicine,  Zhejiang  University,  Hangzhou, China}

\maketitle              
\begin{abstract}
Multi-phase computed tomography (CT) images provide crucial complementary information for accurate liver tumor segmentation (LiTS). State-of-the-art multi-phase LiTS methods usually fused cross-phase features through phase-weighted summation or channel-attention based concatenation. However, these methods ignored the spatial (pixel-wise) relationships between different phases, hence leading to insufficient feature integration. In addition, the performance of existing methods remains subject to the uncertainty in segmentation, which is particularly acute in tumor boundary regions. In this work, we propose a novel LiTS method to adequately aggregate multi-phase information and refine uncertain region segmentation. To this end, we introduce a spatial aggregation module (SAM), which encourages per-pixel interactions between different phases, to make full use of cross-phase information. Moreover, we devise an uncertain region inpainting module (URIM) to refine uncertain pixels using neighboring discriminative features. Experiments on an in-house multi-phase CT dataset of focal liver lesions (MPCT-FLLs) demonstrate that our method achieves promising liver tumor segmentation and outperforms state-of-the-arts.

\end{abstract}
\section{Introduction}
Liver cancer is one of the leading causes of cancer-induced death, which poses a serious risk to human health~\cite{el2020epidemiology}. Accurate liver tumor segmentation (LiTS) is a vital prerequisite for liver cancer diagnosis and treatment, which helps to increase the five-year survival rate. Most existing LiTS solutions~\cite{christ2016automatic,han2017automatic,li2018h,seo2019modified,zhang2019light} tended to use single-phase computed tomography (CT) images to segment liver tumors. However, these methods usually produce unsatisfactory results due to inherent challenges in medical images (e.g., low contrast and fuzzy tumor boundaries). Alternatively, segmentation relying on multi-phase images could assimilate complementary information from different phases, which is helpful to probe the complete morphology of tumors.

In clinical practice, contrast enhanced CT (CECT) images of different phases present distinct liver tumor morphology and gray scales. Generally, tumors may be not salient in one phase but show clear outlines in another phase. Therefore, making good use of the complementary inter-phase information could effectively improve the segmentation results. In view of this, several methods were proposed to explore this issue, which can be classified into three categories according to multi-phase feature fusion strategies: input-level fusion (ILF) methods~\cite{ouhmich2019liver}, decision-level fusion (DLF) methods~\cite{sun2017automatic,ouhmich2019liver,raju2020co} and feature-level fusion (FLF) methods~\cite{wu2019hepatic,xu2021pa}. Among these methods, FLF methods were demonstrated to achieve the best performance since they exploited multi-level cross-phase features. For instance, Wu et al.~\cite{wu2019hepatic} proposed an MW-UNet, which integrated different phases by weighting their features from hidden layers of U-Net~\cite{ronneberger2015u} using trainable coefficients. Xu et al.~\cite{xu2021pa} proposed a ResNet~\cite{he2016deep} based PA-ResSeg to re-weight features of different phases using channel-attention mechanism~\cite{hu2018squeeze,fu2019dual}. However, known FLF methods merely focused on phase-wise or channel-wise inter-phase relationships, but neglected the pixel-wise correspondence between different phases, thus leading to redundancy and low efficiency in information aggregation. The insufficient feature fusion may even bring in interference factors at spatial positions. Moreover, like other segmentation tasks, the performance of existing multi-phase LiTS methods suffer from uncertain region segmentation. That is, the segmentation results usually present some blurry or ambiguous regions (especially in tumor boundaries). This problem is mainly caused by (1) high-frequency information loss during down- and up-sampling operations and (2)low contrast between tumors and surroundings.

Motivated by these observations, in this work, we propose a novel method to segment liver tumors from multi-phase CECT images. In overall, our method exploits complementary information from arterial (ART) phase images to facilitate LiTS in portal venous (PV) phase images. We boost the segmentation performance by introducing pixel-wise inter-phase feature fusion and uncertain region refinement. Specifically, to ensure sufficient multi-phase information aggregation, we devise a spatial aggregation module (SAM). The proposed SAM module mines macro and local inter-phase relationships and yields a pixel-wise response map for each phase. Afterwards, multi-phase features are modulated and fused pixel-by-pixel according to the response maps. Besides, we devise an uncertain region inpainting module (URIM) to refine uncertain regions and obtain fine segmentation. The key idea of the URIM module is to employ confident pixels (with high confidence in segmentation scores) to inpaint surrounding uncertain pixels. To that end, a local-confidence convolution (LC-Conv) operation is introduced to make uncertain pixels absorb neighboring discriminative features. After several LC-Conv operations, the adjusted features are adopted to do the final prediction. Comprehensive experiments on an in-house multi-phase CT dataset of focal liver lesions (MPCT-FLLs) demonstrate that our method achieves accurate liver tumor segmentation and is superior to state-of-the-arts.

Our main contributions are: (1) we devise a spatial aggregation module to ensure sufficient inter-phase interactions. The module extracts macro and local inter-phase relationships, thereby modulating each pixel with a response value; (2) we devise an uncertain region inpainting module to refine uncertain and blurry regions, which particularly helps to obtain fine-grained tumor boundary segmentation; (3) we validate our method on the multi-phase MPCT-FLLs dataset. 

\begin{figure}[t]
\centering
\includegraphics[width=0.9\textwidth]{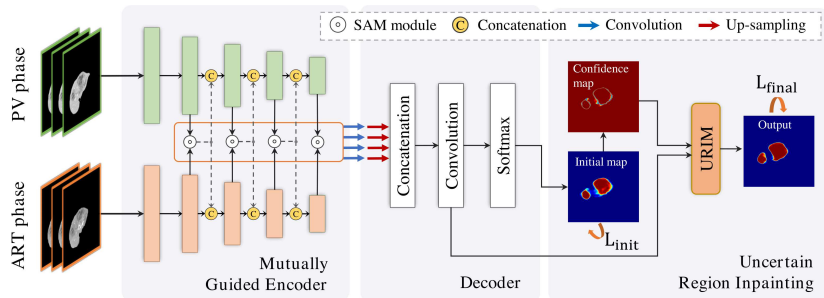}
\caption{Illustration of our proposed multi-phase liver tumor segmentation network.
}
\label{fig_overview}
\end{figure}

\section{Method}
In overall, our network takes both PV- and ART-slices as inputs, and produces tumor segmentation of the primary PV phase. Fig.~\ref{fig_overview} illustrates the overview of the proposed network, which mainly comprises three parts. 

The mutually guided encoder part takes ResNeXt-50~\cite{xie2017aggregated} as the backbone. It uses two Siamese streams, i.e., PV-stream and ART-stream, to extract phase-specific features. Convolution blocks of the two streams are denoted as $\mathrm{B^{(i)}_{PV}}$ and $\rm B^{(i)}_{ART}$ $(\rm i\in\{1,2,3,4,5\}$). To integrate cross-phase information, features from $\mathrm{B^{(i)}_{PV}}$ and $\mathrm{B^{(i)}_{ART}}$ ($\mathrm{ i\in\{2,3,4,5\}}$) are aggregated through SAMs in a bi-directional manner. By doing so, the two streams provide information to assist each other, thus mutually guiding their feature extraction.

The decoder part takes four-level aggregated features from the encoder as input, and yields initial probability maps. To incorporate multi-level features, all inputs are up-sampled using bilinear interpolation, and fused through concatenation and convolutions. 

The uncertain region inpainting part on top of the decoder aims to to refine the uncertain regions in the initial maps.
Intuitively, it employs confident pixels to inpaint neighboring uncertain pixels. To achieve this, uncertain pixels are allowed to absorb surrounding discriminative features using a proposed local-confidence convolution (LC-Conv) operation. The refined features are adopted for the final prediction.

\subsection{Spatial Aggregation Module}
State-of-the-art multi-phase LiTS methods ignored feature fusion at spatial positions. This may lead to redundancy and low efficiency in information integration. Hence, we propose the spatial aggregation module (SAM) to ensure sufficient cross-phase feature fusion by weighting each pixel.

\begin{figure}[t]
\centering
\includegraphics[width=\textwidth]{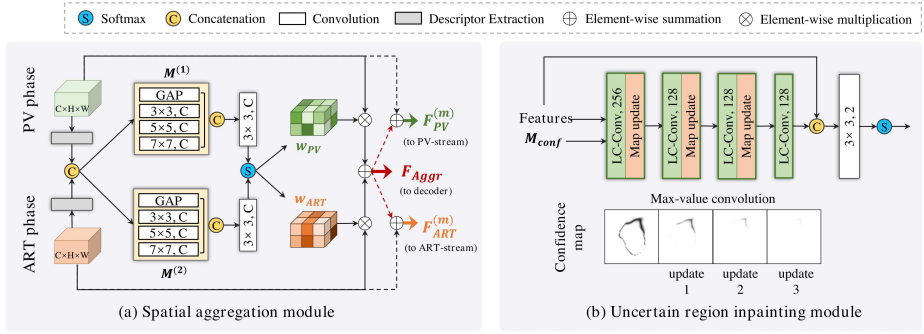}
\caption{Detailed structures of the proposed SAM module (a) and URIM module (b).}
\label{fig_modules}
\end{figure}

Fig.~\ref{fig_modules} (a) shows the detailed structure of the proposed SAM. Having two input feature maps $\mathrm{F_{PV}\in \mathbb{R}^{C\times H\times W}}$ (from PV-stream) and $\mathrm{{F}_{ART}\in\mathbb{R}^{C\times H\times W}}$ (from ART-stream), the SAM module calculates two pixel-wise response maps, which are denoted as $\mathrm{w_{PV}\in\mathbb{R}^{C\times H\times W}}$ and $\mathrm{w_{ART}\in\mathbb{R}^{C\times H\times W}}$, to modulate $\mathrm{{F}_{PV}}$ and $\mathrm{{F}_{ART}}$, respectively. Accordingly, the overall cross-phase feature aggregation can be formulated as:
\begin{equation}
        \rm {F}_{Aggr}=w_{PV}\otimes  {F}_{PV}+w_{ART}\otimes  {F}_{ART}
    \label{eq_aggregation}
\end{equation}
where $\mathrm{{F}_{Aggr}\in\mathbb{R}^{C\times H\times W}}$ is aggregated features; $\otimes $ is element-wise multiplication. 

How to obtain appropriate response maps is the key point of the SAM module. Concretely, the SAM module first extracts efficient descriptors of input features to reduce dimensions and preserve informative characteristics. To do so, we apply average-pooling and max-pooling operations to inputs along the channel direction~\cite{2018CBAM}. The obtained descriptors are denoted as $\mathrm{{F}^{'}_{PV}\in\mathbb{R}^{2\times H\times W}}$ and $\mathrm{{F}^{'}_{ART}\in\mathbb{R}^{2\times H\times W}}$, respectively.
Then, SAM module learns two mapping functions $\mathrm{M^{(1)}}$ and $\rm M^{(2)}$ to model local and global inter-phase complementary relationships from the feature descriptors. Specifically, $\rm M^{(1)}$ and $\rm M^{(2)}$ are built on a pyramid convolution structure (see Fig.~\ref{fig_modules}(a)), i.e., a global average pooling (GAP) layer and a $7\times 7$ convolutional layer are applied to distill global correspondence; two convolutional layers (with kernel sizes of $3\times 3$ and $5\times 5$) are used to capture local inter-phase details. The outputs of $\rm M^{(1)}$ and $\rm M^{(2)}$ are adopted to yield two initial response maps $\mathrm{w^{(0)}_{PV}\in\mathbb{R}^{C\times H\times W}}$ and $\mathrm{w^{(0)}_{ART}\in\mathbb{R}^{C\times H\times W}}$ through concatenation and $3\times3$ convolution (note that we up-sample the output of the GAP layer to $\rm H \times W$ before the concatenation).
The ultimate response maps are obtained by normalizing $\rm w^{(0)}_{PV}$ and $\rm w^{(0)}_{ART}$ through a softmax layer, which ensures $\rm w^{(c,h,w)}_{PV}+w^{(c,h,w)}_{ART}=1$.

So far, the aggregated features $\rm {F}_{Aggr}$ can be calculated using Eq.~\ref{eq_aggregation} and will be fed into the decoder for tumor region prediction. 
Besides, we feed the modulated phase-specific features $\rm {F}^{(m)}_{PV}\in\mathbb{R}^{C\times H\times W}$ and $\rm {F}^{(m)}_{ART}\in\mathbb{R}^{C\times H\times W}$ to PV- and ART-stream to mutually guide their feature extraction, where $\rm {F}^{(m)}_{PV}$ and $\rm {F}^{(m)}_{ART}$ are obtained by:
\begin{equation}
        \rm {F}^{(m)}_{PV}=(F_{PV}+ {F}_{Aggr})/2, \ 
        {F}^{(m)}_{ART}=(F_{ART}+ {F}_{Aggr})/2
\end{equation}

\subsection{Uncertain Region Inpainting Module}
\label{sec_URIM}
The decoding stage took four-level aggregated features from $\mathrm{B^{(i)}_{PV}}$ and $\mathrm{B^{(i)}_{ART}}$ ($\mathrm{ i\in\{2,3,4,5\}}$) to predict preliminary probability maps. However, the initial results usually present some blurry and uncertain regions.
Accordingly, we propose an uncertain region inpainting module (URIM) to refine the ambiguous regions (especially the tumor boundaries). The core idea of our URIM is leveraging pixels with confident classification scores to inpaint neighboring uncertain pixels.

\noindent
\textbf{Confidence Map Calculation.}
Inspired by Liang's work~\cite{liang2020video}, we derive the concept of the confidence map.
Let $\rm S_{i}\in\mathbb{R}^{1\times H\times W} (i\in [1, 2])$ denote the initial segmentation maps. $\rm S_{i}$ is the probability of each pixel $p$ belonging to class $i$ (liver tumor or background), where $\rm \sum_{i=1}^{2}{S_{i}(p)=1}$. Thus, the classification confidence of each pixel can be represented by the confidence map $\rm M_{conf}\in \mathbb{R}^{1\times H\times W}$:
\begin{equation}
       \rm
       M_{conf}=1-exp(1-S^{max}/S^{min})
\end{equation}
where $\rm S^{max}$ represents the largest score of each pixel in initial maps and $\rm S^{min}$ represents the smallest score of each pixel. $\rm M_{conf}$ ranges in $[0,1)$, and a larger value in $\rm M_{conf}$ means the higher confidence. 

\noindent
\textbf{Local-confidence Convolutions.}
Uncertain pixels usually have indistinguishable features, thus it is hard to identify their classes. Intuitively, if we can let uncertain pixels assimilate discriminative features from neighboring confident pixels, the classification of these uncertain pixels may become easier. To achieve this, we propose the local-confidence convolution (LC-Conv) operation, which is formulated as:
\begin{equation}
       \rm
       x^{'}=(W^{T}(X\otimes M_{conf}))/sum(M_{conf})+b
\end{equation}
where $\rm X$ denotes the input features in the current sliding window; $\rm x^{'}$ denotes the refined features; $\rm M_{conf}$ denotes the pixel-wise confidence map; $\rm W$ denotes the weights of the convolution filter and $b$ denotes the bias. The scaling factor $\rm 1/sum(M_{conf})$ is used to regularize the effect of the confidence map within different sliding windows. During each convolution operation, LC-Conv emphasizes discriminative features and suppresses uncertain features. With this mechanism, pixels with higher confidence in the neighboring window contribute more to the filtering result, thereby making uncertain pixels receive surrounding distinguishable features. 
After each LC-Conv operation, $\rm M_{conf}$ is updated through a $3\times3$ max-value convolutional layer.

Fig.~\ref{fig_modules}(b) presents the detailed structure of the URIM module, which consists of four LC-Conv layers with $\rm 3\times3$ kernel. The URIM module takes $\rm M_{conf}$ and decision features from the decoder (feature maps before the softmax layer) as inputs, and yields the refined prediction. During the refinement stage, uncertain pixels gradually absorb more distant confident features while the uncertain regions shrink, which works like the image inpainting~\cite{liu2018image,yu2019free}. At last, we concatenate the refined features and input features to predict the final result.

\subsection{Loss Function and Training Strategy}
\label{sec_loss}
Our loss function comprises two cross-entropy losses (see Fig.~\ref{fig_overview}), i.e., the $L_{init}$ between initial segmentation and ground truths, and $L_{final}$ between final predictions and ground truths. $L_{init}$ and $L_{final}$ contribute equally to the total loss. Our method is implemented based on PyTorch 1.5.0~\cite{paszke2017automatic} and trained on a NVIDIA GTX 2080 ti GPU (12 GB). We use the SGD optimizer to train our network with an initial learning rate of $5 \times 10^{-4}$, which is divided by 10 every 50 epoches.

\section{Materials and Experiments}
\textbf{Dataset.}
We evaluate our method on an in-house multi-phase CT dataset of focal liver lesions (MPCT-FLLs). The dataset contains 121 multi-phase CT cases of five typical liver tumor types with liver and tumor delineations (including 36 cases of cysts, 20 cases of focal nodular hyperplasia (FNH), 25 cases of hemangiomas (HEM), 26 cases of hepatocellular carcinoma (HCC) and 14 cases of metastasis (METS)). The image size is $512\times512$, the slice thickness is 0.5 mm or 0.7 mm and the inter-plane resolution varies from $\rm 0.52\times 0.52 mm^2$ to $\rm 0.86\times 0.86 mm^2$.
To validate our models, we adopt the five-fold cross-validation technique. Accordingly, 121 cases are randomly divided into five mutually-exclusive subsets. All the quantitative results are averaged on five testing sets.

\noindent \textbf{Pre-processing.}
All the images are truncated into the range of [-70, 180] hounsfield unit to eliminate unrelated tissues. Besides, to avoid false-positives outside liver regions, we train a simple ResUNet~\cite{han2017automatic} to segment livers. Each input of our network contains three adjacent slices masked by corresponding liver masks. Besides, we follow Xu's method~\cite{xu2021pa} to register multi-phase inputs, which simply aligned multi-phase tumor volumes according to the tumor center voxels. In the training phase, we augment the data by randomly shifting, rotating and scaling to prevent potential over-fitting problem.

\begin{figure}[t]
\centering
\includegraphics[width=\textwidth]{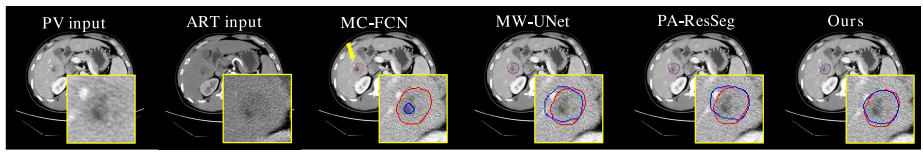}
\caption{Visual comparison between four different segmentation methods. Ground truths are delineated in red and the corresponding predictions are delineated in blue.
}
\label{fig_visual_compare}
\end{figure}

\noindent \textbf{Evaluation Metrics.} 
To quantitatively measure the performance, six common metrics are used, i.e., dice per case (DPC), dice global (DG), volumetric overlap error (VOE), relative volume difference (RVD), average symmetric surface distance (ASSD) and root mean square symmetric surface distance (RMSD). The higher DPC and DG scores mean the better segmentation results. For the rest four evaluation metrics, the smaller the absolute value is, the better the results.

\begin{table}[t]
\renewcommand\tabcolsep{3pt}
\caption{Quantitative comparison between our method and state-of-the-arts.}
\label{tb_compare}
\centering
\begin{tabular}{lcccccc}
\toprule
Methods &
  \begin{tabular}[c]{@{}c@{}}DPC \\(\%)\end{tabular} &
  \begin{tabular}[c]{@{}c@{}}DP\\ (\%)\end{tabular} &
  \begin{tabular}[c]{@{}c@{}}VOE\\ (\%)\end{tabular} &
  \begin{tabular}[c]{@{}c@{}}RVD \\(\%)\end{tabular} &
  \begin{tabular}[c]{@{}c@{}}ASD\\ (mm)\end{tabular} &
  \begin{tabular}[c]{@{}c@{}}RSMD\\ (mm)\end{tabular} \\ \midrule
MC-FCN~\cite{sun2017automatic}    & 51.80          & 71.63          & 59.30          & -5.92         & 27.26         & 30.45         \\
MW-UNet~\cite{wu2019hepatic}   & 73.37          & 86.34          & 37.10          & 10.11         & 10.73         & 18.16         \\
PA-ResSeg~\cite{xu2021pa} & 77.26          & 86.21          & 33.46          & \textbf{4.46} & 3.71          & 7.82          \\
\textbf{Ours}      & \textbf{80.12} & \textbf{86.51} & \textbf{30.08} & -5.28         & \textbf{2.81} & \textbf{5.47} \\ \midrule
ILF & 73.19 & 83.13 & 37.6 & -17.69 & 14.2 & 18.15 \\
DLF~\cite{sun2017automatic} & 75.87 & 85.36 & 35.14 & -5.73 & 4.45 & 9.51 \\
MW~\cite{wu2019hepatic} &77.39 & 85.52 & 33.07 & 5.43 & 6.31 & 8.89 \\
PA~\cite{xu2021pa} & 78.24 & 86.29 & 32.46 & 6.01 & 3.39 & 6.90 \\
\textbf{SAM (Ours)} & \textbf{80.12} & \textbf{86.51} & \textbf{30.08} & \textbf{-5.28}  & \textbf{2.81} & \textbf{5.47} \\
\bottomrule
\end{tabular}
\end{table}

\begin{table}[t]
\renewcommand\tabcolsep{3pt}
\caption{Quantitative comparison (in DPC (\%)) regarding to different tumor types.}
\label{tb_tumor_types}
\centering
\begin{tabular}{lcccccc}
\toprule
Methods   & Cyst           & FNH            & HCC            & HEM            & METS           & Average        \\ \midrule
MC-FCN    & 67.82          & 21.58          & 46.98          & 51.59          & 61.07          & 51.80          \\
MW-UNet   & 83.07          & 64.18          & 64.80          & 71.42          & 79.35          & 73.37          \\
PA-ResSeg & 85.70          & \textbf{73.13} & 71.89          & 74.63          & 79.64          & 77.26          \\
\textbf{Ours}      & \textbf{90.87} & 66.08          & \textbf{79.11} & \textbf{75.42} & \textbf{83.27} & \textbf{80.12}\\ \bottomrule
\end{tabular}
\end{table}

\noindent \textbf{Comparison with State-of-the-arts.}
To validate the effectiveness of the proposed method, we compare it with three state-of-the-art multi-phase LiTS methods: (1) MC-FCN~\cite{sun2017automatic}, which simply concatenated multi-phase features before the classification layer of FCN~\cite{long2015fully}; (2) MW-UNet~\cite{wu2019hepatic}, which trained a specific weight value for each phase at multiple layers of the U-Net; (3) PA-ResSeg~\cite{xu2021pa}, which incorporated channel-attention mechanism to re-weight each channels of multi-phase features at specific layers of the ResNet. 

Fig.~\ref{fig_visual_compare} depicts the visual example of the comparison experiments. It is observed that MC-FCN produces poor results as it rudely concatenated decision-level features from different phases. These raw multi-phase features may bring in conflicts or interference factors, thus leading to the bad performance. MW-UNet and PA-ResSeg produces better results and are able to capture the rough tumor shape. However, the pixel-wise segmentation, especially on the tumor boundaries, is not satisfactory. The reason is that both of the methods neglected the spatial-wise feature fusion, thereby aggregating multi-phase features insufficiently. Besides, they did not provide any strategies to handle the uncertainty problem, which made it hard to do boundary pixel classification. In contrast, our method encourages per-pixel inter-phase interactions and incorporates uncertain region inpainting mechanism, which is demonstrated to achieve the best value in DPC, DP, VOE, ASD and RSMD scores (see Table.~\ref{tb_compare}).

To analyze the performance of our method on specific tumor types, we divide the testing sets into five subsets according to tumor categories. As seen in Table.~\ref{tb_tumor_types}, our method boosts prominent performance gains on most types of tumors compared to other methods, and achieves the top performance on the average.

Further, to validate our proposed cross-phase fusion strategy (the SAM module), we compare it with other four fusion methods: (1) Input-level fusion (ILF) strategy, which concatenates multi-phase images in the inputs; (2) Decision-level fusion (DLF) strategy used in MC-FCN~\cite{sun2017automatic}; (3) Modality weighting (MW) strategy adopted in MW-UNet~\cite{wu2019hepatic} and (4) Phase Attention (PA) strategy adopted in PA-ResSeg~\cite{xu2021pa}. To ensure the fairness of comparison, all the fusion modules are plugged into the proposed network except that the feature aggregation module is replaced. As shown in Table.~\ref{tb_compare}, our SAM strategy makes better use of multi-phase information and outperforms other fusion strategies.

\begin{table}[t]
\renewcommand\tabcolsep{3pt}
\caption{Quantitative results for ablation studies.}
\label{tb_ablation}
\centering
\begin{tabular}{llllcccccc}
\toprule
SP &
  MP &
  SAM &
  URIM &
  \multicolumn{1}{c}{\begin{tabular}[c]{@{}c@{}}DPC\\ (\%)\end{tabular}} &
  \multicolumn{1}{c}{\begin{tabular}[c]{@{}c@{}}DG\\ (\%)\end{tabular}} &
  \multicolumn{1}{c}{\begin{tabular}[c]{@{}c@{}}VOE\\ (\%)\end{tabular}} &
  \multicolumn{1}{c}{\begin{tabular}[c]{@{}c@{}}RVD\\ (\%)\end{tabular}} &
  \multicolumn{1}{c}{\begin{tabular}[c]{@{}c@{}}ASD\\ (mm)\end{tabular}} &
  \multicolumn{1}{c}{\begin{tabular}[c]{@{}c@{}}RMSD\\ (mm)\end{tabular}}\\ \midrule
 \checkmark&  &  &  & 68.85 & 83.31 & 40.07 & -19.22 & 21.53& 27.73\\
 \checkmark&  \checkmark&  &  & 75.31 & 85.13 & 35.97 & -5.89 &5.00&10.05 \\
\checkmark &  \checkmark&  \checkmark&  & 79.10 & 86.12 & 31.22 & -5.51&3.44&6.46   \\
 \checkmark&  \checkmark&  \checkmark&  \checkmark& \textbf{80.12} & \textbf{86.51} & \textbf{30.08} & \textbf{-5.28}&\textbf{2.81}&\textbf{5.47} \\ \bottomrule
\end{tabular}
\end{table}

\noindent \textbf{Ablation Study.}
To validate the effectiveness of each component in our method, we start from a single-phase (PV phase) ResNeXt-50 network and gradually add the modules. Table.~\ref{tb_ablation} summaries the quantitative results, in which SP means single-phase segmentation and MP means multi-phase segmentation by simply adding PV- and ART-features at four convolution blocks (we denote this fusion strategy as FLF-add). It is seen that adding multi-phase information donates a performance boost of +6.46\% in DPC; employing SAM modules to fuse features improves the performance by +3.79\% in DPC; refining uncertain regions via the URIM module contributes a performance gain of +1.02\% in DPC.

\noindent \textbf{Robustness Validation.}
Our SAM and URIM modules could be plugged into various multi-phase segmentation networks. To validate their robustness on different backbones, we replace our backbone with U-Net~\cite{ronneberger2015u} and ResNet-50~\cite{he2016deep}. The DPC gains of SAM and URIM modules on multi-phase U-Net, ResNet-50 and ResNeXt-50 (adopting FLF-add strategy) are 5.28\%, 4.50\% and 4.81\%, respectively. It is demonstrated that the SAM and URIM modules can improve the segmentation performance stably in different networks.

\section{Discussion and Conclusions}
In this work, we propose a novel multi-phase network to segment liver tumors from PV- and ART-phase images. In our network, we devise a spatial aggregation module to take full advantage of multi-phase information by modulating each pixel. We also devise an uncertain region inpainting module to handle uncertainty in tumor boundary segmentation. Experiments on the MPCT-FLLs dataset demonstrates the superiority of our method against state-of-the-arts. In the future, we may focus on multi-phase segmentation problem on more than two phases. In this situation, we may use the SAM module (with more parameters) to generate the response map for each phase. This issue will be more challenging since we need to consider the memory consumption of the network.
%
%
%
\bibliographystyle{splncs04}
\bibliography{paper198}

\begin{thebibliography}{10}
\providecommand{\url}[1]{\texttt{#1}}
\providecommand{\urlprefix}{URL }
\providecommand{\doi}[1]{https://doi.org/#1}

\bibitem{christ2016automatic}
Christ, P.F., Elshaer, M.E.A., Ettlinger, F., Tatavarty, S., Bickel, M., Bilic,
  P., Rempfler, M., Armbruster, M., Hofmann, F., D’Anastasi, M., et~al.:
  Automatic liver and lesion segmentation in {CT} using cascaded fully
  convolutional neural networks and {3D} conditional random fields. In:
  International Conference on Medical Image Computing and Computer-Assisted
  Intervention. pp. 415--423. Springer (2016)

\bibitem{el2020epidemiology}
El-Serag, H.B.: Epidemiology of hepatocellular carcinoma. The Liver: Biology
  and Pathobiology pp. 758--772 (2020)

\bibitem{fu2019dual}
Fu, J., Liu, J., Tian, H., Li, Y., Bao, Y., Fang, Z., Lu, H.: Dual attention
  network for scene segmentation. In: Proceedings of the IEEE/CVF Conference on
  Computer Vision and Pattern Recognition. pp. 3146--3154 (2019)

\bibitem{han2017automatic}
Han, X.: Automatic liver lesion segmentation using a deep convolutional neural
  network method. arXiv preprint arXiv:1704.07239  (2017)

\bibitem{he2016deep}
He, K., Zhang, X., Ren, S., Sun, J.: Deep residual learning for image
  recognition. In: Proceedings of the IEEE Conference on Computer Vision and
  Pattern Recognition. pp. 770--778 (2016)

\bibitem{hu2018squeeze}
Hu, J., Shen, L., Sun, G.: Squeeze-and-excitation networks. In: Proceedings of
  the IEEE conference on computer vision and pattern recognition. pp.
  7132--7141 (2018)

\bibitem{li2018h}
Li, X., Chen, H., Qi, X., Dou, Q., Fu, C.W., Heng, P.A.: H-denseunet: hybrid
  densely connected unet for liver and tumor segmentation from {CT} volumes.
  IEEE Transactions on Medical Imaging  \textbf{37}(12),  2663--2674 (2018)

\bibitem{liang2020video}
Liang, Y., Li, X., Jafari, N., Chen, Q.: Video object segmentation with
  adaptive feature bank and uncertain-region refinement. In: Advances in neural
  information processing systems (NeurIPS) (2020)

\bibitem{liu2018image}
Liu, G., Reda, F.A., Shih, K.J., Wang, T.C., Tao, A., Catanzaro, B.: Image
  inpainting for irregular holes using partial convolutions (2018)

\bibitem{long2015fully}
Long, J., Shelhamer, E., Darrell, T.: Fully convolutional networks for semantic
  segmentation. In: Proceedings of the {IEEE} Conference on Computer Vision and
  Pattern Recognition. pp. 3431--3440 (2015)

\bibitem{ouhmich2019liver}
Ouhmich, F., Agnus, V., Noblet, V., Heitz, F., Pessaux, P.: Liver tissue
  segmentation in multiphase ct scans using cascaded convolutional neural
  networks. International journal of computer assisted radiology and surgery
  \textbf{14}(8),  1275--1284 (2019)

\bibitem{paszke2017automatic}
Paszke, A., Gross, S., Chintala, S., Chanan, G., Yang, E., DeVito, Z., Lin, Z.,
  Desmaison, A., Antiga, L., Lerer, A.: Automatic differentiation in pytorch
  (2017)

\bibitem{raju2020co}
Raju, A., Cheng, C.T., Huo, Y., Cai, J., Huang, J., Xiao, J., Lu, L., Liao, C.,
  Harrison, A.P.: Co-heterogeneous and adaptive segmentation from multi-source
  and multi-phase ct imaging data: A study on pathological liver and lesion
  segmentation. In: European Conference on Computer Vision. pp. 448--465.
  Springer (2020)

\bibitem{ronneberger2015u}
Ronneberger, O., Fischer, P., Brox, T.: U-net: Convolutional networks for
  biomedical image segmentation. In: International Conference on Medical Image
  Computing and Computer-assisted Intervention. pp. 234--241. Springer (2015)

\bibitem{seo2019modified}
Seo, H., Huang, C., Bassenne, M., Xiao, R., Xing, L.: Modified u-net (mu-net)
  with incorporation of object-dependent high level features for improved liver
  and liver-tumor segmentation in {CT} images. IEEE Transactions on Medical
  Imaging  \textbf{39}(5),  1316--1325 (2019)

\bibitem{sun2017automatic}
Sun, C., Guo, S., Zhang, H., Li, J., Chen, M., Ma, S., Jin, L., Liu, X., Li,
  X., Qian, X.: Automatic segmentation of liver tumors from multiphase
  contrast-enhanced ct images based on fcns. Artificial intelligence in
  medicine  \textbf{83},  58--66 (2017)

\bibitem{2018CBAM}
Woo, S., Park, J., Lee, J.Y., Kweon, I.S.: Cbam: Convolutional block attention
  module  (2018)

\bibitem{wu2019hepatic}
Wu, Y., Zhou, Q., Hu, H., Rong, G., Li, Y., Wang, S.: Hepatic lesion
  segmentation by combining plain and contrast-enhanced ct images with modality
  weighted u-net. In: 2019 IEEE International Conference on Image Processing
  (ICIP). pp. 255--259. IEEE (2019)

\bibitem{xie2017aggregated}
Xie, S., Girshick, R., Doll{\'a}r, P., Tu, Z., He, K.: Aggregated residual
  transformations for deep neural networks. In: Proceedings of the IEEE
  conference on computer vision and pattern recognition. pp. 1492--1500 (2017)

\bibitem{xu2021pa}
Xu, Y., Cai, M., Lin, L., Zhang, Y., Hu, H., Peng, Z., Zhang, Q., Chen, Q.,
  Mao, X., Iwamoto, Y., et~al.: Pa-resseg: A phase attention residual network
  for liver tumor segmentation from multiphase ct images. Medical Physics
  (2021)

\bibitem{yu2019free}
Yu, J., Lin, Z., Yang, J., Shen, X., Lu, X., Huang, T.S.: Free-form image
  inpainting with gated convolution. In: Proceedings of the IEEE/CVF
  International Conference on Computer Vision. pp. 4471--4480 (2019)

\bibitem{zhang2019light}
Zhang, J., Xie, Y., Zhang, P., Chen, H., Xia, Y., Shen, C.: Light-weight hybrid
  convolutional network for liver tumor segmentation. In: IJCAI. pp. 4271--4277
  (2019)

\end{thebibliography}
\end{document}